\documentclass[conference]{IEEEtran}
\IEEEoverridecommandlockouts
\usepackage{cite}
\usepackage{amsmath,amssymb,amsfonts}
\usepackage{bm}
\usepackage{algorithmic}
\usepackage{graphicx}
\usepackage{textcomp}
\usepackage{xcolor}
\usepackage{algorithm}
\usepackage{subcaption}
\usepackage{float}
\usepackage{array}
\usepackage{stfloats}
\usepackage{url}
\def\BibTeX{{\rm B\kern-.05em{\sc i\kern-.025em b}\kern-.08em
    T\kern-.1667em\lower.7ex\hbox{E}\kern-.125emX}}
\begin{document}

\title{Joint SIM Configuration and Power Allocation for Stacked Intelligent Metasurface-assisted MU-MISO Systems with TD3 
}
 \author{\IEEEauthorblockN{Xiaolei Yang\IEEEauthorrefmark{1}, Jiayi Zhang\IEEEauthorrefmark{1}, Enyu Shi\IEEEauthorrefmark{1}, Ziheng Liu\IEEEauthorrefmark{1}, Jun Liu\IEEEauthorrefmark{2}, Kang Zheng\IEEEauthorrefmark{2}\IEEEauthorrefmark{3} and Bo Ai\IEEEauthorrefmark{1}}\\
 \IEEEauthorblockA{
 \IEEEauthorrefmark{1}School of Electronic and Information Engineering, Beijing Jiaotong University, Beijing 100044, China\\
 \IEEEauthorrefmark{2}China Mobile Zijin Innovation Institute, Nanjing 210031, China\\
 \IEEEauthorrefmark{3}National Mobile Communications Research Laboratory, Southeast University, Nanjing 210096, China\\
 }}

\maketitle

\begin{abstract}
The stacked intelligent metasurface (SIM) emerges as an innovative technology with the ability to directly manipulate electromagnetic (EM) wave signals, drawing parallels to the operational principles of artificial neural networks (ANN).  Leveraging its structure for direct EM signal processing alongside its low-power consumption, SIM holds promise for enhancing system performance within wireless communication systems. In this paper, we focus on SIM-assisted multi-user multi-input and single-output (MU-MISO) system downlink scenarios in the transmitter. We proposed a joint optimization method for SIM phase shift configuration and antenna power allocation based on the twin delayed deep deterministic policy gradient (TD3) algorithm to efficiently improve the sum rate. The results show that the proposed algorithm outperforms both deep deterministic policy gradient (DDPG) and alternating optimization (AO) algorithms. Furthermore, increasing the number of meta-atoms per layer of the SIM is always beneficial. However, continuously increasing the number of layers of SIM does not lead to sustained performance improvement. 

\end{abstract}

\section{Introduction}

The performance of reconfigurable intelligent surfaces (RIS) in wireless communication systems is constrained by its single-layer structure \cite{9614196} and discrete phase shift capabilities \cite{10158356}. 
In contrast, stacked intelligent metasurface (SIM) has captured widespread interest with its ability to directly manipulate electromagnetic wave signals, drawing parallels to the operational principles of artificial neural networks (ANN) \cite{doi:10.1126/science.aat8084,liu2022programmable}. 
By configuring the phase shift of each meta-atom per layer in the SIM for complex signal processing tasks like beamforming in wireless communication system, the antennas can just transmit the analog data stream converted by the low-resolution digital to analog covert (DAC) without additional signal processing circuitry \cite{10158690,an2023stacked}. 
Furthermore, the stacked structure and continuous phase shift configuration of meta-atoms offer superior beamforming accuracy \cite{10379500}, enhancing signal quality, mitigating path loss, and reducing user interference. However, it's important to note that the multilayer structure of SIM, while enabling complex electromagnetic signal processing, also introduces substantial computational requirements for configuring the phase shifts of each meta-atom per layer.

Several prior works have focused on the downlink of SIM-assisted MU-MISO communication systems. 
\textit{An et al.} \cite{10279173} first introduced SIM to a MU-MISO communication system. By proposing an alternative optimization (AO) algorithm, they jointly optimized the power allocation and SIM-assisted beamforming to improve the system's sum rate and mitigate user interference. Extensively, in \cite{liu2024drlbased}, \textit{Liu et al.} leveraged deep deterministic policy gradient (DDPG) algorithm to address the joint power allocation and beamforming problem efficiently. The numerical results demonstrate the superior performance of the deep reinforcement learning approach over traditional algorithms. 
However, while DDPG showed promise, the complexities of the SIM structure and wireless channel state suggest a need for more advanced optimization methods.

In this context, the twin delayed deep deterministic policy gradient (TD3) algorithm presents a promising approach. As an extension of DDPG, TD3 introduces twin critic networks and a delayed updates strategy, effectively relieving the overestimation bias and high variance estimation. Besides, TD3 exploits policy noise and policy smoothing strategies to facilitate the agent's exploration ability. 
Considering the features of the downlink of SIM-assisted systems such as numerous parameters of the phase shifts, complex state of the wireless channel and stacked structure of the SIM, TD3 can solve the problem more efficiently for its exceptional convergence speed, stability, and generalization ability.

Motivated by the aforementioned observations, in this work, we propose a joint SIM phase shift configuration and power allocation method based on the twin delayed deep deterministic policy
gradient (TD3) algorithm for improving the sum rate of SIM-assisted systems. Our main contributions are as follows:
\begin{itemize} 
\item We investigate the system performance of the SIM-assisted MU-MISO system. Specifically, we formulate the sum rate maximization problem through joint consideration of SIM phase shift configuration and AP power allocation.

\item We propose an advanced TD3-based algorithm in tackling this complex non-convex optimization problem. Due to the stability and convergence properties exhibited by TD3, experiment results demonstrate the superior performance of our proposed algorithm over AO \cite{10279173} and DDPG \cite{liu2024drlbased} approaches.  

\item We investigate the impact of SIM’s structure on system performance. In particular, increasing the number of meta-atoms per layer is always beneficial. However, continuously increasing the number of layers of SIM does not lead to sustained performance improvement. 
\end{itemize} 

\section{System Model}

\begin{figure}[t]
  \centering
  \includegraphics[width=0.45\textwidth]{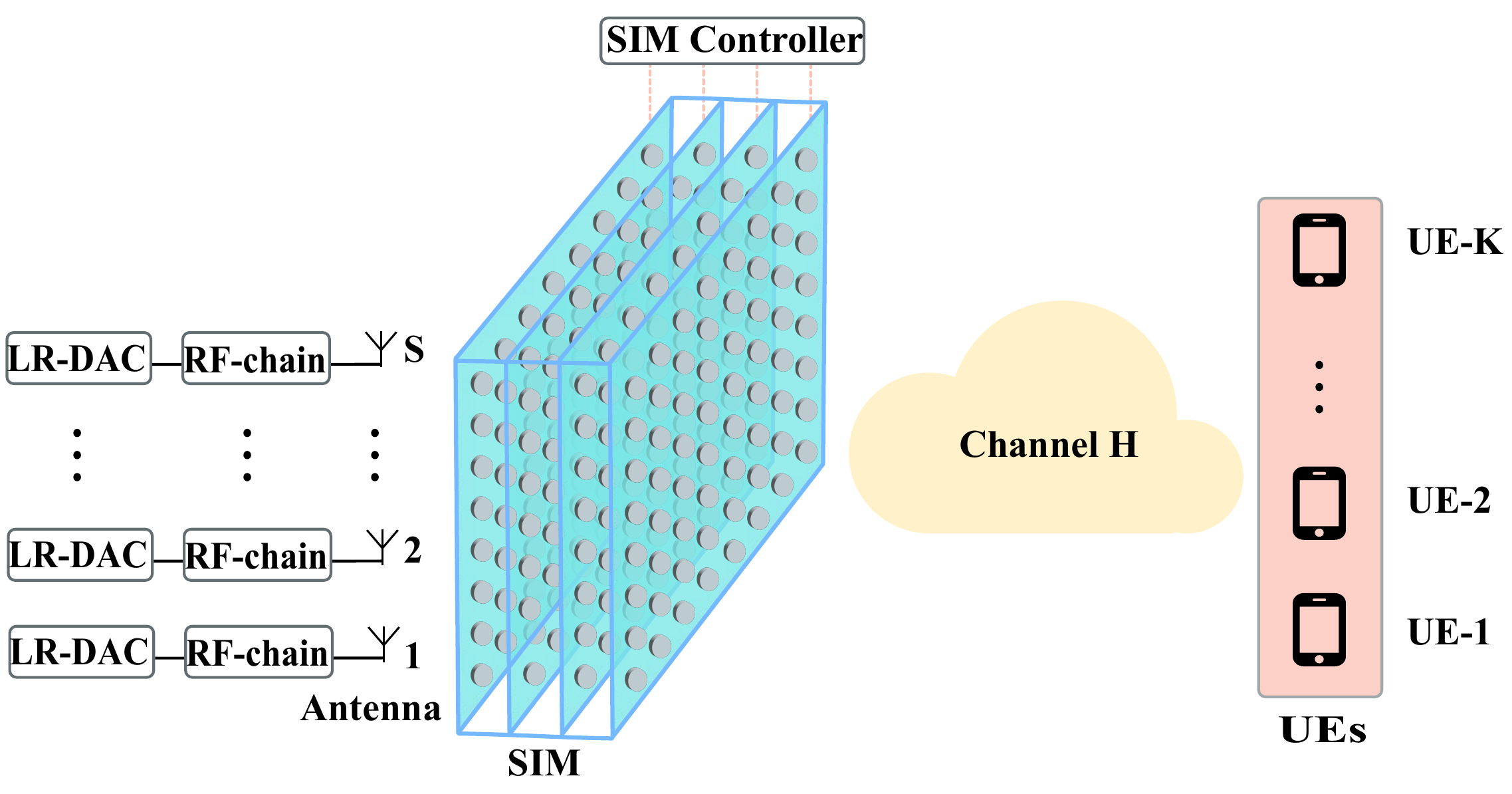}
  \caption{A SIM-assisted MU-MISO downlink system}
  \label{fig:sim}\vspace{-0.4cm}
\end{figure}

In Fig.~\ref{fig:sim}, we consider a SIM-assisted MU-MISO system comprising $K$ single-antenna users (UEs) and a single base station (BS) with $S$ antennas and one SIM device. The SIM consists of $L$ metasurface layers, each with $M$ meta-atoms, enabling precise phase shift adjustments for downlink beamforming in the EM wave domain.

\subsection{SIM Model}

Let ${\mathcal{M}}=\left\{ {1,2, \cdots M} \right\}$ and ${\mathcal{L}}=\left\{ {1,2, \cdots, L} \right\}$ represent the number of meta-atoms of each SIM of the SIM and the number of layers. 
Due to the meta-atoms capability to change signal's phase when the EM wave passes through it, let $\varphi _m^l = {e^{j\theta _m^l}}$ denote the phase shift coefficient of the $m$-th meta-atom on the $l$-th metasurface layer, while $\varphi _m^l$ represents the phase shift generated by the signal passing through the meta-atom.
In this way, the diagonal phase shift matrix of the whole metasurface for the $l$-th layer is written as ${\mathbf{\Phi}^l} = {\rm{diag}}\left( {\varphi _1^l,\varphi _2^l, \cdots,\varphi _M^l} \right) \in \mathbb{C} {^{M \times M}},\forall l \in \mathcal{L}$. 

The propagation of EM waves between adjacent metasurfaces obeys the Rayleigh-Sommerfeld diffraction theory \cite{doi:10.1126/science.aat8084}, \cite{10279173}. Hence, the propagation coefficient between the $\tilde m$-th meta-atom on the ($l-1$)-th metasurface layer and the $m$-th meta-atom on the $l$ metasurface layer is defined by 
\begin{equation}
    w_{m,\tilde m}^l \!=\! \frac{{{d_x}{d_y}{z_s}}}{{d_{m,\tilde m}^l}}\left( {\frac{1}{{2\pi d_{m,\tilde m}^l}} \!-\! j\frac{1}{\lambda }} \right){e^{j2\pi d_{m,\tilde m}^l/\lambda }},\forall l \ne 1,l \in {\mathcal{L}},
  \label{eq:w}
\end{equation}
where $\lambda$ is the wavelength, ${d_{m,\tilde m}^l}$ represents the propagation length between the meta-atoms, $z_{s}$ represents the length between interlayer space, and ${d_x} \times {d_y}$ is the space of each meta-atom. The propagation matrix from the ($l-1$)-th to the $l$-th can be denoted as $\mathbf{W}^l \in \mathbb{C} {^{M \times M}},\forall l \ne 1,l \in {\mathcal{L}}$. 
The propagation vector from the $s$-th antenna to the SIM's first metasurface layer $\mathbf{w}_{s}^1$ can be calculated by replacing $\mathit d_{m,\tilde m}^l$ of (\ref{eq:w}) by the length between $s$-th antenna $d_{m,s}^{1}$ and the $m$-th meta-atoms on the first metasurface.

Therefore, the SIM-assisted wave domain beamforming matrix is given as
\begin{equation}
    \mathbf{G} = {\mathbf{\Phi} ^L}{\mathbf{W}^L}{\mathbf{\Phi} ^{L - 1}}{\mathbf{W}^{L - 1}} \cdots {\mathbf{\Phi} ^2}{\mathbf{W}^2}{\mathbf{\Phi} ^1} \in \mathbb{C} {^{M \times M}}
    \label{eq:G}.
\end{equation}

\subsection{MU-MISO Communication System Model}

Let ${\mathcal{S}}=\left\{ {1,2, \cdots  S} \right\}$ and ${\mathcal{K}}=\left\{ {1,2, \cdots,K} \right\}$ represent the number of antennas at the BS and the number of single-antenna users. Assuming that these $S$ antennas can implement transmit $S$ data streams corresponding to a specific number of UEs, $S$ and $K$ must satisfy the condition $S = kK, k \in \mathbb{Z}{^+}$. In our work, we assume that $S$ is equal to $K$, which means that every antenna from the BS can transmit and receive the individual user data streams directly. 

In a SIM-assisted wireless communication system, the processed signals are effectively emitted from closely spaced meta-atoms on the final metasurface of the SIM to the users. Thus, we adopted a spatially correlated Rayleigh fading channel model for accurately modeling the system. The channel matrix from the $k$-th user is written as $\mathbf{h}_{k}^{H} \in \mathbb{C}^{1 \times M}, \forall k \in K$, and $\mathbf{h}_{k} \sim \mathcal{CN} (\mathbf{0},\mathcal{\beta}_k \mathbf{R})$, where $\beta_k$ is the path loss of the link between the $k$-th UE and the BS, and $\mathbf{R}\in \mathbb{C}^{M \times M}$ is the spatial correlation matrix of the last SIM metasurface. 
What's more, the path loss from the $k$-th UE to the BS is given as $\beta_{k} = \beta_0 (\mathit{d} / \mathit{d}_{0})$, where $\beta_0$ denotes the path loss at a baseline distance of $\mathit{d}_{0}$. In an isotropic scattering environment with uniformly distributed multipath components, the ($m,m'$)-th entry of $\mathbf{R}$ is $\mathbf{R}_{m,m'}= {\rm{sinc}}(2\mathit{d}_{m,m'}/{\lambda})$, where $\rm{sinc(\mathit{x})} =\rm{sin(\mathit{\pi x})/(\mathit{\pi x})}$ and $\mathit{d}_{m,m'}$ represents the length between adjacent meta-atoms.

From the perspective of UEs, the received signal can be written as:
\begin{align}\label{eq:y_K}
{y_k}&= \mathbf{h}_{\text{k}}^{\text{H}}\mathbf{G}\sum\limits_{j = 1}^K {\mathbf{w}_j^1\sqrt {{p_j}} {x_j}}  + {n_k} \notag\\     &=\mathbf{h}_{\text{k}}^{\text{H}}\mathbf{G}\mathbf{w}_k^1\sqrt {{p_k}} {x_k} + \sum\limits_{j \ne k}^K {\mathbf{h}_{\text{k}}^{\text{H}}\mathbf{G}\mathbf{w}_j^1\sqrt {{p_j}} {x_j}}  + {n_k} 
    ,\forall k \in \mathcal{K},
\end{align}
where $\mathbf{h}_k^H$ is the channel matrix from the meta-atoms in the final layer to $k$-th UE, $\mathbf{G}$ is the wave-based beamforming matrix of the SIM, $\mathbf{w}_{k}^1$ is the propagation vector from $k$-th antenna to the first SIM layer and $n_k \sim \mathcal{CN} (0,\sigma_k^2)$ is the additive white Gaussian noise with variance $\sigma_k^2$. Let donate $\mathbf{p}={\rm{diag}}([p_1, p_2, \cdots, p_K])$ as the power allocation vector and $p_k$ represents the power allocation to the $k$-th UE where $p_k \geq 0$ and ${\textstyle \sum_{k = 1}^K {{p_k}} \leqslant {P_T}}$. The $x_k$ is the information symbol received by the $k$-th UE and the vector $\mathbf{x}$ represents the data streams transmitted to all users characterized by zero mean and unit variance.

Therefore, the received signal-to-interference-plus-noise-ratio (SINR) at the $\mathit{k}$-th user is given by \cite{10167480}
\begin{equation}
    \label{eq:gama}
    {\gamma _k} = \frac{|\mathbf{h}_{\text{k}}^{\text{H}}\mathbf{G}\mathbf{w}_k^1|^2 {p_k}}{{\textstyle \sum_{j \ne k}^K }{|\mathbf{h}_{\text{k}}^{\text{H}}\mathbf{G}\mathbf{w}_j^1|^2 {p_j} + \sigma _k^2} }
    , \forall k \in \mathcal{K}.
\end{equation}

\subsection{Problem Formulation}

In our work, we are interested in utilizing SIM devices to improve the Quality-of-Service (QoS) in the MU-MISO system. Therefore our optimization goal is to jointly design the SIM phase shifts matrix $\bf{\Phi}$ and the antenna power allocation vector $\bf{p}$ for relieving the multiuser interference and maximizing the system's sum-rate performance.
Assuming that the perfect CSI is known by the BS, we can formulate the joint optimization problem as follows:
\begin{align}
    \label{eq:max}
        (P1) \quad &  \mathop{\max}\limits_{\bm{\Phi},\mathbf{p}}\mathbf{R}=\sum\limits_{k=1}^K{\mathrm{\log}_2 (1+\gamma_k})  
      \\
    		\textbf{s.t.} \quad 
    		& \sum_{k=1}^{K}p_k \leq P_t , \tag{5a}\\
            & p_k \geq 0 , \forall k \in \mathcal{K}, \tag{5b}   \\
            & \theta_m^l \in \left [ 0,2\pi  \right ) ,\forall m \in \mathcal{M},\forall l \in \mathcal{L}. \tag{5c}   
\end{align}
The problem stated in (\ref{eq:max}) is complex and challenging due to its non-convex objective function and constraints. Traditional mathematical methods would require an exhaustive search for an optimal solution, especially in large-scale networks, making it impractical. Therefore, in this paper, instead of solving the difficult optimization problem directly, we use a novel DRL method to formulate and find the solution for the sum rate optimization problem based on the TD3 algorithm.

\begin{figure}[t]
  \centering
  \includegraphics[width=0.49\textwidth]{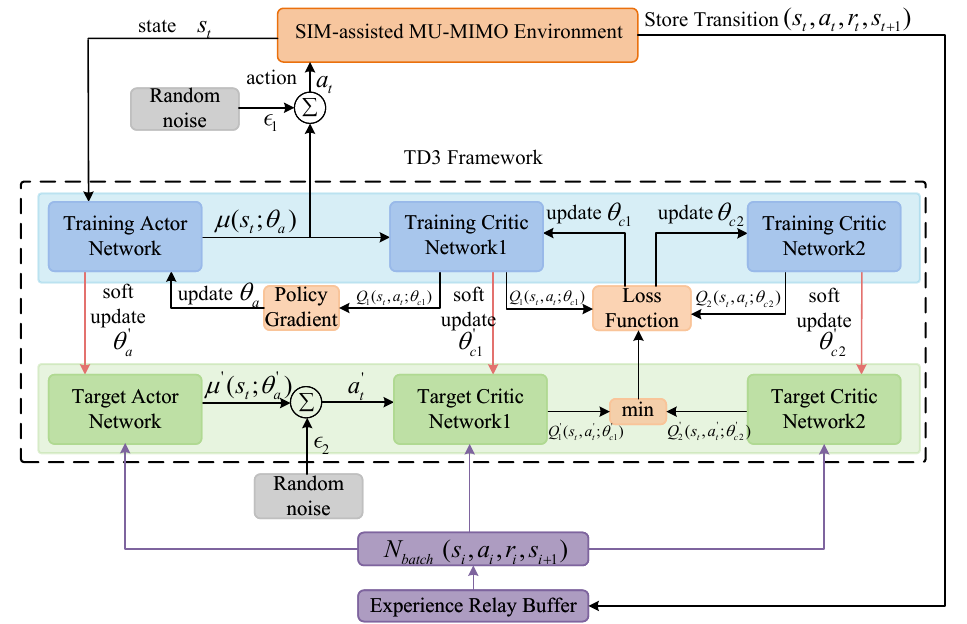}
  \caption{Proposed TD3-based framework}
  \label{fig:dlr}
  \vspace{-0.4cm}
\end{figure}

\section{Proposed TD3-based Optimization}
In this section, we first give an insight into the TD3 algorithm framework and explain why the TD3 algorithm is well-suited for improving the sum rate in complex SIM-assisted communication scenarios. We then present the specifics of our joint SIM phase shift configuration and power allocation method based on the TD3 algorithm for improving the SIM-assisted system's sum rate.
\begin{algorithm}[t]
    \caption{Joint optimization algorithm based on TD3}
    \label{alg:td3}
    \begin{algorithmic}[1]
        \STATE $\bf{Input}$: channel matrix $\mathbf{h}_k^{H}, \forall k \in \mathcal{K}$, SIM propagation matrix $\mathbf{W}$, replay buffer, batch size $N_{batch}$.
        \STATE $\bf{Output}$: Optimal action $a$, Optimal $\mathbf{R}$.
        \STATE $\bf{Initialization}$: Randomly initialize the training actor network $\mu(s_t;\theta_a)$ with the weights $\theta_a$, the training critic network $Q_{1},Q_{2}$ with the weights $\theta_{c1}$,$\theta_{c2}$ and the target networks $Q_{1}^{'}$,$Q_{2}^{'}$ with the weights $\theta_{c1}^{'} \gets \theta_{c1}$,$\theta_{c2}^{'} \gets \theta_{c2}$.
        \FOR{each episode}
            \STATE Get initial state $s_1$.
            \FOR{each step, $t = 1,2,\cdots,T$ }
                \STATE Select action $a_t = \mu (s_t;\theta_a) + \epsilon_1,\epsilon_1 \sim \mathcal{N}(0,\sigma_1^2)$.
                \STATE Calculate the beamforming matrix based on (\ref{eq:G}), obtain the reward from (\ref{eq:max}), and get the next state $s_{(t+1)}$.
                \STATE Store the experience $(s_t, a_t, r_t, s_{(t+1)})$ in replay buffer.
                \STATE Sample minibatches of size $N_{batch}$ from replay buffer randomly.
                \STATE Update the training critic network parameters according to the loss function (\ref{eq:loss}).
                \STATE Update the training actor network parameters according to the policy gradient according to (\ref{eq:pg}).
                \STATE Update the target network parameters according to (\ref{eq:tc}), (\ref{eq:ta}).
                \STATE Set state = next state.
            \ENDFOR
        \ENDFOR
    \end{algorithmic}
    
\end{algorithm}

\subsection{Twin Delayed Deep Deterministic Policy Gradient}
DDPG is a classical DRL algorithm that has been widely used for optimization tasks due to its ability to handle continuous action spaces \cite{9769985}. TD3, as a recent extension of DDPG, comprises six networks depicted in Fig.~\ref{fig:dlr}. The training-actor network, denoted as $\mu(s_t;\theta_a)$ with $\theta_a$ representing the network parameters, provides an approximate policy for the agent and generates the action $a_t$ in state $s_t$. The two training-critic networks, $Q_1(s_t,a_t;\theta_{c1})$ and $Q_2(s_t,a_t;\theta_{c2})$ with parameters $\theta_{c1}$ and $\theta_{c2}$, estimate the action value function $Q = \min\{Q_1(s_t,a_t;\theta_{c1}),Q_2(s_t,a_t;\theta_{c2})\}$ based on actions from the training-actor network. The target-actor network, $\mu^{'}(s_{(t+1)} ; \theta^{'}_a)$, is designed to generate the target action $a_t^{'}$ for training the target-critic networks. Similarly, the two target-critic networks, $\mathit{Q}_{1}^{'}(s_{(t+1)},a_t^{'};\theta_{c1}^{'})$ and $Q_2^{'}(s_{(t+1)},a_t^{'};\theta_{c2}^{'})$, generate the target Q-value for training the training-actor and training-critic networks.

In network parameter updating, the loss function typically quantifies the disparity between the predicted value by the neural network and the ground truth. However, in reinforcement learning, the objective is to approximate the optimal Q-value function, for which the ground truth is unknown. To address this issue, TD3 employs two target-critic networks to obtain the actual target value \cite{10146001}.
Therefore, the actual target value and the loss function for the training-critic networks are given as
\begin{align}
    &y = r + \beta \min \{ Q_{1}^{'}(s_{(t+1)},a_t^{'};\theta_{c1}^{'}), Q_{2}^{'}(s_{(t+1)},a_t^{'};\theta_{c2}^{'})\}, \\
    &\mathit{loss}(\theta) = ({y - \min\{Q_{1}(s_t,a_t;\theta_{c1}),Q_{2}(s_t,a_t;\theta_{c2})\}})^2, \label{eq:loss}
\end{align}
where $\beta \in (0,1] $ is the discount rate. 
The update for the training critic network is based on the gradient of the loss function (\ref{eq:loss}). The training-actor network parameters updating is defined through the policy gradient theorem \cite{9864655}. And the target network parameters by the soft updating policy. These parameters updating equations are written as
\begin{align}
    &\theta _c \gets \theta _c -\alpha_c \nabla_{\theta _c}loss(\theta_c),\\
    &\theta _a \gets \theta _a -\alpha_a \nabla_{\mu}\mathit{Q}(s_t,\mu(s_t;\theta _a);\theta _c)\nabla _{\theta_a}\mu(s_t;\theta _a), \label{eq:pg}\\
    &\theta_c ^{'}\gets \tau_c \theta_c +(1-\tau_c ) \theta_c ^{'} \label{eq:tc},\\
    &\theta_a ^{'}\gets \tau_a \theta_a +(1-\tau_a ) \theta_a ^{'} \label{eq:ta} , 
\end{align}
where $\alpha_a, \alpha_c$ represent the learning rate for the training actor and critic network and $\tau_a, \tau_c$ represent the learning rates for updating the target networks.

The detailed steps of the TD3 algorithm are shown in  Algorithm \ref{alg:td3} above.
Notably, TD3 introduces twin critic networks and a delayed updates strategy which relieves the overestimation bias and high variance estimation present in DDPG by smoothing target Q-values and updating target networks with a delay \cite{pmlr-v80-fujimoto18a}. Moreover, TD3 exploits policy random noise and target policy smoothing strategies to facilitate the agent's exploration ability and improve training stability. Considering the outstanding advantages in terms of convergence speed, stability, and generalization ability, TD3 performs better in high-dimension continuous action space which is more suitable for optimization in SIM-assisted communication system.  

\begin{figure}[t]
  \centering
  \includegraphics[width=0.4\textwidth]{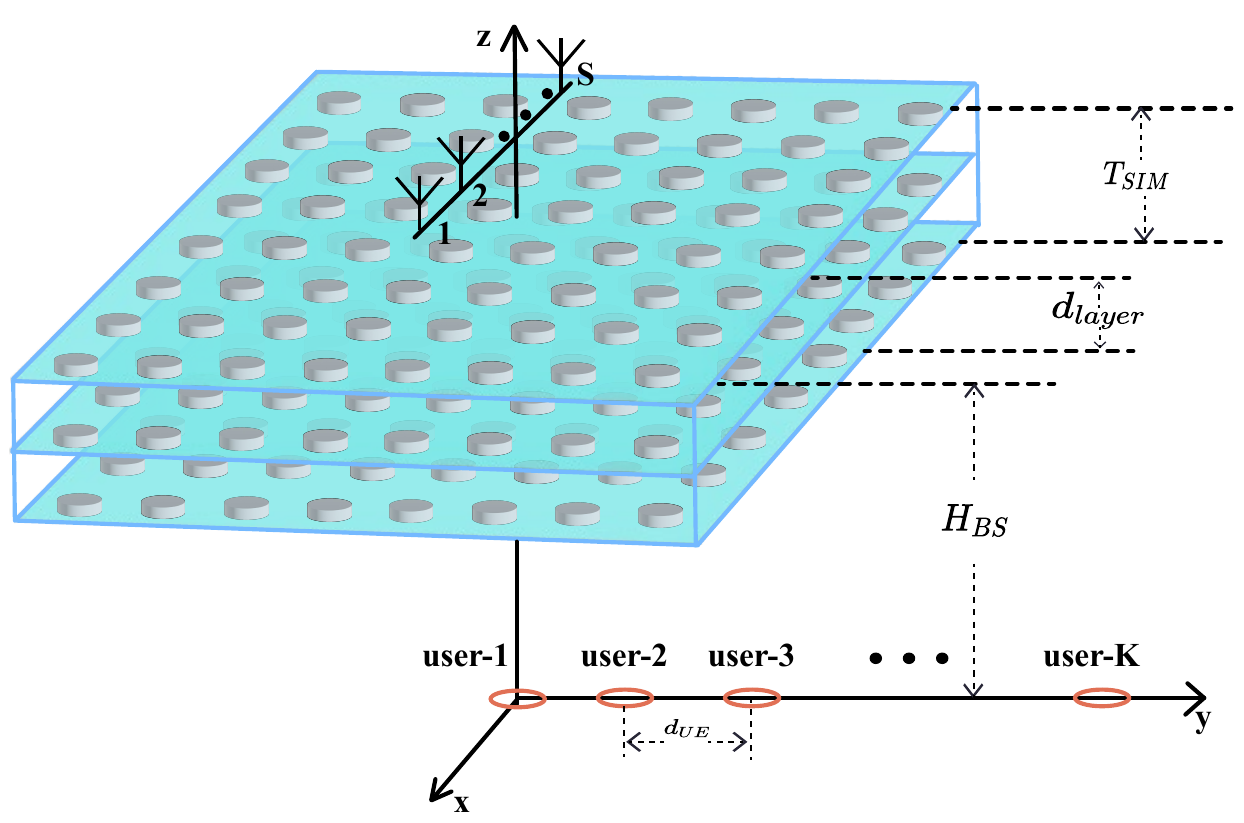}
  \caption{Simulation setup of the SIM-assisted MU-MISO communication system.}
  \label{fig:sim_2}\vspace{-0.4cm}
\end{figure}

\subsection{Basic Elements of DRL}
TD3 has the same elements such as actions, states, rewards, and other elements like conventional reinforcement learning algorithms. The detailed design of our TD3 optimization is as follows:

\subsubsection{Action $a$}
It represents a set of actions chosen by the agent following a policy $\pi$ and the actions are the matrix $\mathbf{\Phi}$ and $\mathbf{P}$ in our work. Due to the deep neural network can only take real numbers as input, the matrix $\mathbf{\Phi}$ is separated as the real part and the imaginary part for proper input. Thus, the action dimension of matrix $\mathbf{\Phi}$ is $2ML$, the dimension of matrix $\mathbf{P}$ is $K$ and the total dimension is $2ML + K$.

\subsubsection{State $s$}
It is the set of observations that are obtained from the environment. In our communication system, the channel matrix $\mathbf{H}$ is the significant information that promotes the agent to understand the conditions of the wireless channel. Besides, the matrix $\mathbf{\Phi}$ and $\mathbf{P}$ have an impact on the SINR and the sum rate of the system. Therefore, the state at the time $t$ , $s_t$ is consist of $\mathbf{\Phi}$,  $\mathbf{P}$ and $\mathbf{H}$ with the dimension of $2ML + K + 2MK$.

\subsubsection{Reward $r$}
The reward is instant feedback from the environment by taking an action in the current state. Based on the reward, the agent can evaluate the action in the current state and learn superior policy.  In our paper, we define the sum rate function $\mathbf{R}$ as the reward.

\subsubsection{Policy $\pi$}
the policy $\pi(s_t; \theta_a)$ represents a guideline for the agent taking actions conditioned on the state $s_t$. 
\subsubsection{Experience}
defined as $(s_t,a_t,r_{t},s_{(t+1)})$.

\subsubsection{Loss Function}
The loss function is discussed in the last subsection.

\section{Numerical Results }

\begin{table}[t]
    \centering
    \caption{TD3 Hyperparameters}
    \label{tab:td3_hyperparameters}
    \begin{tabular}{c c c}
        \hline
        parameter & description & value \\
        \hline
        $\tau_a$  & training actor network learning rate &  0.005 \\
        $\tau_c$   & training critic network learning rate & 0.005  \\
        $\alpha_a$   & target actor network soft update rate & 0.0003 \\
        $\alpha_c$  & target critic network soft update rate & 0.0003 \\
        $\beta$  & discount rate &  0.9 \\
        B  & replay buffer size & 1000000  \\
        $N_{batch}$  & batch size & 256 \\
        $\epsilon_1$ & action explosion noise & 0.02 \\
        $\epsilon_2$  & policy noise power & 0.04 \\
        $N_t$  & the total steps of each episode & 6000 \\
        $N$  & the number of episodes & 100 \\
        \hline
    \end{tabular}
\end{table}
\subsection{Simulation Setup}
According to Fig.~\ref{fig:sim_2}, we assume a SIM-assisted MU-MISO downlink system with the carrier frequency of 28 GHz. The BS is equipped with $S$ antennas aligned parallel to the $x$-axis. The SIM consisting of $L$ layers is in front of these antennas and the metasurfaces are deployed to achieve the SIM wave-based beamforming from the BS. The center of the BS and SIM is on the $z$-axis, and the height from antennas to surface $z=0$ is set to $H_{BS}=10$ m. The length from the first to the last layer of SIM is defined as $T_{SIM} = 5\lambda$, and the length between two layers can be calculated as $d_{layer} = T/L$. Every metasurface has $M$ meta-atoms and we assume that these $M$ atoms are uniformly arranged in an $N$ by $N$ square array on the metasurface where $ M = N^2$. The length and width of the meta-atom are defined as $d_{x}=\lambda/2$ and $d_y=\lambda/2$ \cite{shi2022wireless}, respectively. The $K$ users are arranged along the $y$-axis and the distance between every two users is $d_{UE}=10$ m. The path loss from the $k$-th UE to the BS is given as $\beta_{k} = \beta_0 (d / d_{0})$, and we assume that $d_0 = 1$ m and $ \beta_0 = -30$ dB. The max transmit power is set to $P_t = 10$ dBm and the channel noise power is $\sigma_k^2=-104$ dBm, $\forall k \in \mathcal{K}$.

As to the TD3 network, the actor network is composed of an input layer corresponding to the DRL state space dimension, two fully connected layers with dimensions 400 and 300 utilizing tanh($x$) as the activation function. The critic network has the same architecture as the actor network. TD3 network's hyperparameters are presented in the TABLE \ref{tab:td3_hyperparameters}.

To assess the effectiveness of our TD3-based optimization approach, we compare it against three benchmark schemes: 1) The DDPG method, which employs a standard deep reinforcement learning framework for joint precoding and power allocation, as described in \cite{10279173}. 2) The AO method, which utilizes iterative water-filling (IWF)  for transmit power allocation and gradient ascent for optimizing the SIM phase shift, as outlined in \cite{liu2024drlbased}. 3) The IWF method solely optimizes transmit power allocation using the iterative water-filling algorithm, while the phase shift of the SIM is randomized. These benchmark schemes provide a basis for evaluating the performance and efficacy of our TD3-based optimization strategy.

\subsection{Sum Rate versus the Number of Metasurface Layers}
\begin{figure}[t]
\centering
        \includegraphics[scale=0.5]{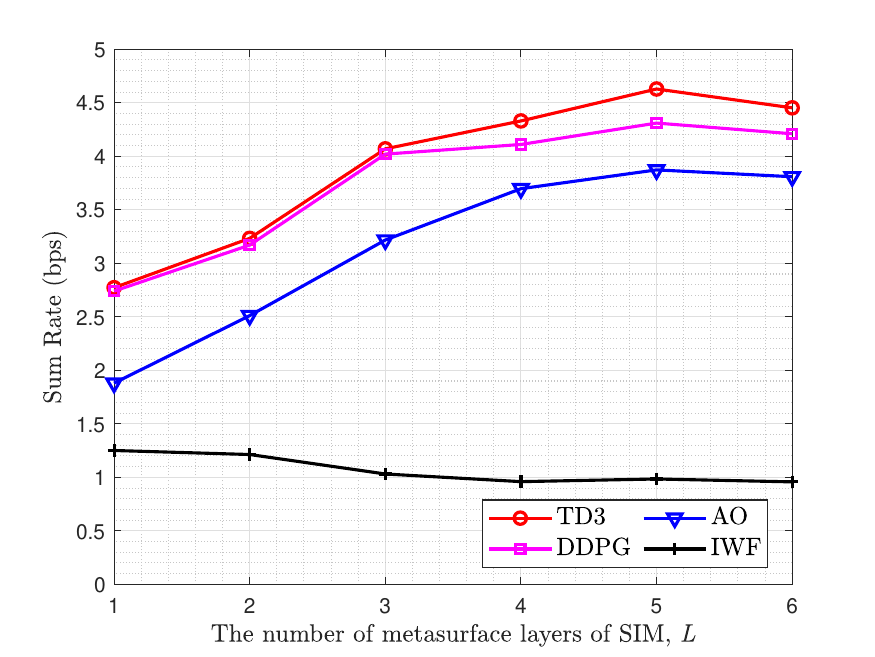}
        \caption{Sum rate $R_{\text{sum}}$ versus the number of SIM layers $L$ ($S$ = 4, $K$ = 4, $M$ = 9, and $P_t$ = 10 dBm).}
        \label{fig:L}\vspace{-0.4cm}
\end{figure}

\begin{figure}[t]
\centering
        \includegraphics[scale=0.5]{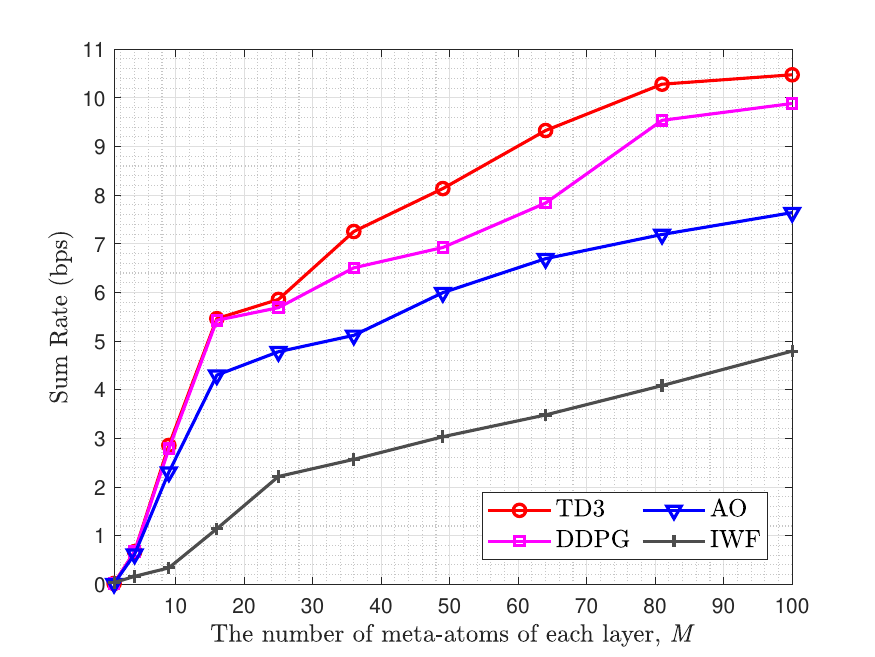}
        \caption{Sum rate $R_{\text{sum}}$ versus the number of meta-atom of each layer $M$ ($L$ = 2, $S$ = 4, $K$ = 4, and $P_t$ = 10 dBm).}
        \label{fig:M}\vspace{-0.4cm}
\end{figure}

Fig.~\ref{fig:L} illustrates the impact of the number of SIM layers on the sum rate. The plot shows the sum rate $R_{\text{sum}}$ plotted against the number of metasurface layers $L$, assuming parameters $S$ = 4, $K$ = 4, $M$ = 9, and $P_T = 10$ dBm. With the layers the number of layers increasing, we can see that the DRL-based algorithms outperform traditional alternating optimization methods. DRL algorithms learn complex, adaptive strategies directly from data and their robustness to various challenges making them highly effective in optimization tasks compared to traditional methods. Besides, the performance of TD3 and DDPG is almost identical when $L$ is very small. However, as $L$ increases, the action space dimension of the agent becomes larger, and the more appropriately designed TD3 algorithm demonstrates superiority over DDPG. Regardless of the method, when the number of layers $L$ is large, the optimization effect of the system will also tend to saturate.

\begin{figure}[t]
\centering
        \includegraphics[scale=0.5]{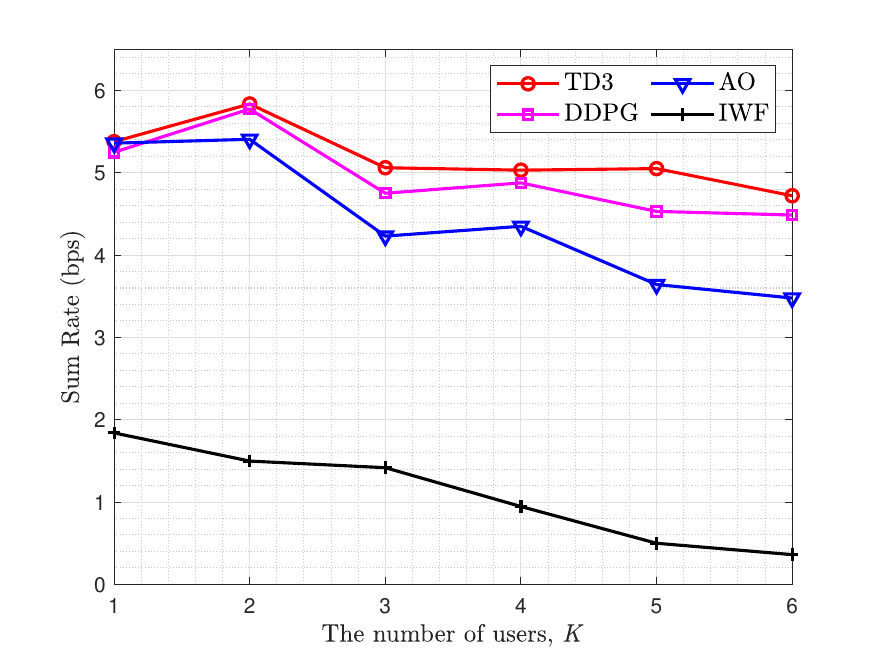}
        \caption{Sum rate $R_{\text{sum}}$ versus the number of users $K$ ($M$ = 9, $S$ = $K$, $L$ = 4, and $P_t$ =10 dBm).}
        \label{fig:K}\vspace{-0.4cm}
\end{figure}

\subsection{Sum Rate versus the Number of Meta-atoms in each Layer}
In Fig.~\ref{fig:M}, as the number of meta-atoms of each metasurface layer increases, the system's rate $R_{\text{sum}}$ also increases. In this scenario, we assume $L$ = 2, $S$ = 4, $K$ = 4 and $P_t$ = 10 dBm.
With an increase in the number of atoms, we can observe that the system sum rate also increases. This implies that more atoms bring about higher-resolution phase shifts leading to higher accuracy in downlink beamforming by increasing the degree of freedom (DoF) of the metasurface \cite{shi2023ris}. However, increasing the number of atoms comes at the cost of algorithm complexity. We can observe that the optimization performance of the AO algorithm is not as effective as that of the DRL algorithm when optimizing a large number of parameters. Furthermore, the efficiency of the TD3 algorithm is higher than DDPG when $M \geq 25$.

\subsection{Sum Rate versus the Number of Users}
According to Fig.~\ref{fig:K}, we see that as the number of users increases, the sum rate decreases with the assumption that $M$ = 9, $S$ = 4, $K$ = 4 and $L$ = 4. The sum rate is declining with the increasing of users although the use of SIM optimization relieves the fact. This is because as the number of users increases, the interference between different users, as shown in (\ref{eq:gama}), also becomes more severe. This also indicates that in SIM-assisted MU-MISO systems, relying solely on the transmitter deployment of SIM cannot completely solve the problem of multi-user interference. In contrast, for the issue of multi-user interference, we can attempt to deploy SIM at the receiver to eliminate the interference, which will also be our future work.

\subsection{Complexity and Convergence Analysis}
The complexity of the DRL algorithm depends on the network structure and the total number of training steps, denoted as $N_T$. The actor network comprises one input layer, two hidden layers, and one output layer with dimensions $D_1$, $L_1$, $L_2$, and $D_2$. The critic network consists of two networks, critic1, and critic2, sharing the same structure with one input layer, two hidden layers, and one output layer having dimensions $D_3$, $L_1$, $L_2$, and 1. As to \cite{zhong2022deep}, our algorithm's complexity is $\mathcal{O}[N_T[D_1L_1+L_1L_2+L_2D_2+2(D_3L_1+L_1L_2+L_2)]]$. The additional complexity of TD3 compared to DDPG arises from introducing two extra critic networks for enhanced stability. 

\begin{figure}[t]
\centering
        \includegraphics[scale=0.5]{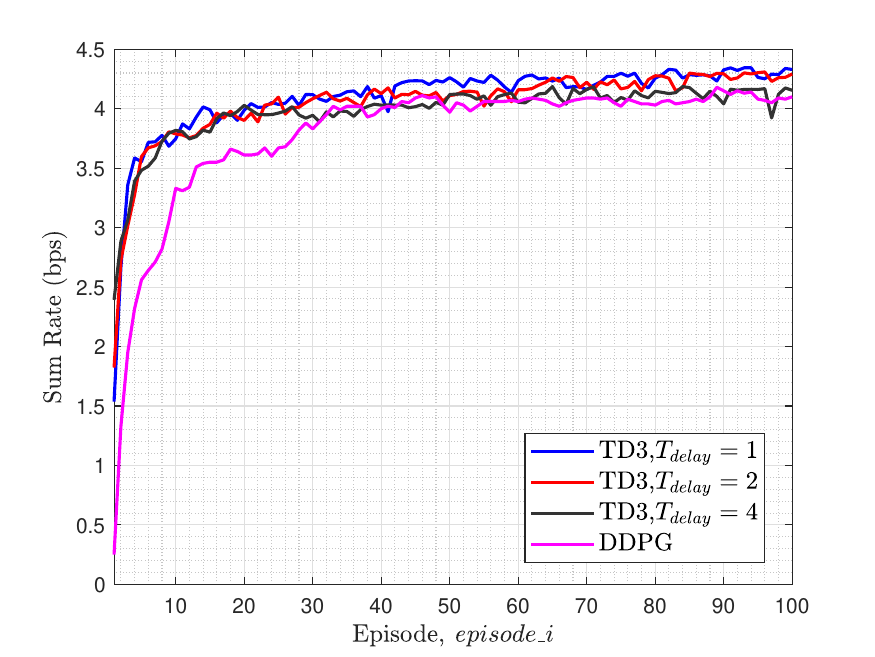}
        \caption{The convergence curve of the sum rate under different delayed update time ($L$ = 5, $S$ = 4, $K$ = 4, $M$ = 9, and $P_t$ = 10 dBm).}
        \label{fig:RL}\vspace{-0.4cm}
\end{figure}

From Fig.~\ref{fig:RL}, we can see the impact of different delayed actor update time. This strategy helps to reduce update frequency, prevent overfitting and oscillation, and improve the algorithm generalization ability. We plotted the training convergence curves for three different delay update times: $T_{delay}$ = 1, $T_{delay}$ = 2, and $T_{delay}$ = 4. Compared with DDPG, we can see the stability that TD3 exhibits. When the delay update time is set too long, we can observe a slowdown in the convergence speed of the algorithm, and the excessively long update interval also leads to the algorithm inability to adapt to environmental changes. When the delayed update time is too short, the algorithm oscillations become particularly severe. 

\section{Conclusions}
In this work, our focus lies on the SIM-assisted MU-MISO communication systems. 
Specifically, we propose a novel method based on the TD3 algorithm for jointly considering SIM phase shift configuration and antenna power allocation. Through experimental comparisons with DDPG and AO methods, we demonstrate the superior performance of the proposed TD3-based algorithm. It is important to show that increasing the number of meta-atoms per layer of the SIM is always beneficial. However, escalating the number of SIM layers does not result in a continual enhancement in performance. 
In the future, our research endeavors will be focused on investigating more efficient methods for SIM configuration within more complex systems, such as massive MIMO systems and cell-free networks.
\bibliographystyle{IEEEtran}
\bibliography{main}

\end{document}